\documentclass{IEEEcsmag}
\newcommand{\maglayout}{}
\usepackage[utf8]{inputenc}
\usepackage[T1]{fontenc} 
\RequirePackage[hyphens]{url}
\usepackage{lscape}
\usepackage[colorlinks,urlcolor=blue,linkcolor=blue,citecolor=blue]{hyperref}
\usepackage{booktabs} 
\usepackage{amsfonts} 
\usepackage{nicefrac} 
\usepackage{microtype}
\usepackage{xcolor}
\usepackage{comment}
\usepackage{multirow}
\usepackage{graphicx}
\usepackage{tabularx}
\usepackage{rotating}
\usepackage{makecell}
\usepackage{tikz}
\usepackage[capitalise,noabbrev,nameinlink]{cleveref}

\usepackage{tcolorbox}

\newtcbox{\trendbox}[1][blue!30]{on line,
    colback=#1, colframe=#1, boxsep=0pt, boxrule=0pt, size=small, arc=1mm,right=-1pt,left=-1pt,top=-1pt, bottom=-1pt}

\newcommand{\harmBox}[1]{\trendbox[gray!30]{#1}}

\usepackage{caption,subcaption}

\definecolor{Orange}{rgb}{1,0.5,0}
\definecolor{Red}{rgb}{1,0,0}
\definecolor{Green}{rgb}{0,0.65,0.5}
\definecolor{Purple}{rgb}{0.75,0,1}
\definecolor{babypink}{rgb}{0.96, 0.76, 0.76}
\definecolor{azure}{rgb}{0,0.49,1}
\definecolor{periwinkle}{rgb}{0.8, 0.8, 1.0}
\definecolor{Pink}{RGB}{255, 102, 204}

\usepackage{upmath}

\renewcommand{\paragraph}[1]{\smallskip \noindent \textit{#1}}

\jvol{XX}
\jnum{XX}
\paper{8}
\jmonth{May/June}
\jname{IT Professional}
\pubyear{2019}

\newcommand\blfootnote[1]{%
  \begingroup
  \renewcommand\thefootnote{}\footnote{#1}%
  \addtocounter{footnote}{-1}%
  \endgroup
}

\RequirePackage[normalem]{ulem}
\RequirePackage{color}\definecolor{RED}{rgb}{1,0,0}\definecolor{BLUE}{rgb}{0,0,1}
 
\providecommand{\DIFaddbegin}{} 
\providecommand{\DIFaddend}{} 
\providecommand{\DIFdelbegin}{} 
\providecommand{\DIFdelend}{} 
\providecommand{\DIFaddbeginFL}{} 
\providecommand{\DIFaddendFL}{} 
\providecommand{\DIFdelbeginFL}{} 
\providecommand{\DIFdelendFL}{} 
\newcommand{\DIFscaledelfig}{0.5}
\RequirePackage{settobox} 
\RequirePackage{letltxmacro} 
\newsavebox{\DIFdelgraphicsbox} 
\newlength{\DIFdelgraphicswidth} 
\newlength{\DIFdelgraphicsheight} 
\LetLtxMacro{\DIFOincludegraphics}{\includegraphics} 
\newcommand{\DIFaddincludegraphics}[2][]{{\color{blue}\fbox{\DIFOincludegraphics[#1]{#2}}}} 
\newcommand{\DIFdelincludegraphics}[2][]{
\sbox{\DIFdelgraphicsbox}{\DIFOincludegraphics[#1]{#2}}
\settoboxwidth{\DIFdelgraphicswidth}{\DIFdelgraphicsbox} 
\settoboxtotalheight{\DIFdelgraphicsheight}{\DIFdelgraphicsbox} 
\scalebox{\DIFscaledelfig}{
\parbox[b]{\DIFdelgraphicswidth}{\usebox{\DIFdelgraphicsbox}\\[-\baselineskip] \rule{\DIFdelgraphicswidth}{0em}}\llap{\resizebox{\DIFdelgraphicswidth}{\DIFdelgraphicsheight}{
\setlength{\unitlength}{\DIFdelgraphicswidth}
\begin{picture}(1,1)
\thicklines\linethickness{2pt} 
{\color[rgb]{1,0,0}\put(0,0){\framebox(1,1){}}}
{\color[rgb]{1,0,0}\put(0,0){\line( 1,1){1}}}
{\color[rgb]{1,0,0}\put(0,1){\line(1,-1){1}}}
\end{picture}
}\hspace*{3pt}}} 
} 
\LetLtxMacro{\DIFOaddbegin}{\DIFaddbegin} 
\LetLtxMacro{\DIFOaddend}{\DIFaddend} 
\LetLtxMacro{\DIFOdelbegin}{\DIFdelbegin} 
\LetLtxMacro{\DIFOdelend}{\DIFdelend} 
\DeclareRobustCommand{\DIFaddbegin}{\DIFOaddbegin \let\includegraphics\DIFaddincludegraphics} 
\DeclareRobustCommand{\DIFaddend}{\DIFOaddend \let\includegraphics\DIFOincludegraphics} 
\DeclareRobustCommand{\DIFdelbegin}{\DIFOdelbegin \let\includegraphics\DIFdelincludegraphics} 
\DeclareRobustCommand{\DIFdelend}{\DIFOaddend \let\includegraphics\DIFOincludegraphics} 
\LetLtxMacro{\DIFOaddbeginFL}{\DIFaddbeginFL} 
\LetLtxMacro{\DIFOaddendFL}{\DIFaddendFL} 
\LetLtxMacro{\DIFOdelbeginFL}{\DIFdelbeginFL} 
\LetLtxMacro{\DIFOdelendFL}{\DIFdelendFL} 
\DeclareRobustCommand{\DIFaddbeginFL}{\DIFOaddbeginFL \let\includegraphics\DIFaddincludegraphics} 
\DeclareRobustCommand{\DIFaddendFL}{\DIFOaddendFL \let\includegraphics\DIFOincludegraphics} 
\DeclareRobustCommand{\DIFdelbeginFL}{\DIFOdelbeginFL \let\includegraphics\DIFdelincludegraphics} 
\DeclareRobustCommand{\DIFdelendFL}{\DIFOaddendFL \let\includegraphics\DIFOincludegraphics} 
\RequirePackage{listings} 
\RequirePackage{color} 
\lstdefinelanguage{DIFcode}{ 
  moredelim=[il][\color{red}\sout]{\%DIF\ <\ }, 
  moredelim=[il][\color{blue}\uwave]{\%DIF\ >\ } 
} 
\lstdefinestyle{DIFverbatimstyle}{ 
	language=DIFcode, 
	basicstyle=\ttfamily, 
	columns=fullflexible, 
	keepspaces=true 
} 
\lstnewenvironment{DIFverbatim}{\lstset{style=DIFverbatimstyle}}{} 
\lstnewenvironment{DIFverbatim*}{\lstset{style=DIFverbatimstyle,showspaces=true}}{}

\begin{document}

\sptitle{Department}
\editor{Editor}

\title{Safer Digital Intimacy For Sex Workers And Beyond: A Technical Research Agenda}

\author{Vaughn Hamilton\IEEEauthorrefmark{1}}
\affil{Max Planck Institute for Software Systems}

\author{Gabriel Kaptchuk\IEEEauthorrefmark{1}}
\affil{Boston University}

\author{Allison McDonald\IEEEauthorrefmark{1}}
\affil{Boston University}

\author{Elissa M. Redmiles\IEEEauthorrefmark{1}}
\affil{Georgetown University}

\markboth{Department Head}{Safer Digital Intimacy: A Technical Research Agenda}

\begin{abstract}
Many people engage in digital intimacy: sex workers, their clients, and people who create and share intimate content recreationally. With this intimacy comes significant security and privacy risk, exacerbated by stigma. In this article, we present a commercial digital intimacy threat model and 10 research directions for safer digital intimacy.
\end{abstract}

\maketitle

\ifdefined\maglayout
\chapteri{P}eople engage in sexual intimacy online in a variety of ways.
\else
\section{Introduction}
People engage in sexual intimacy online in a variety of ways.
\fi
\blfootnote{\IEEEauthorrefmark{1}Following the norm in math and theoretical computer science, authors are listed alphabetically.}
For example, they may send intimate images to each other (sext) or they may work in or consume content produced by the sex industry. Engaging in digital intimacy is common: an estimated 1 in every 200 people has been a sex worker (eg. escorting, stripping, webcamming, sexting, etc.) 
in their lifetime and more than 80\% of adults have sexted.\footnote{The most recent rigorous estimate of sexting behavior was produced in 2015: \url{https://www.apa.org/news/press/releases/2015/08/common-sexting}.}

Many risks plague online sexual intimacy, both as work (sex workers) and recreation (sexters). For example, the theft and resharing of intimate media is a significant and growing form of sexual abuse for everyone engaged in digital intimacy~\cite{eaton20172017}. People must carefully vet those they intend to meet in person, whether from a dating platform or a sex work platform, to prevent stalking, harassment and violence. Online platforms are hostile to both sexual expression and to sex work, regardless of the legal status or nature of the content~\cite{spivsak2021social,barwulor2021disadvantaged,blunt2020erased,blunt2020posting}.

Thus, even though maintaining safety while engaging in digital intimacy is necessary for authentic online engagement, ensuring this safety 
can be challenging. 

In this piece we review \& synthesize ten directions for cryptographic and systems-security research that could improve security and privacy for digital intimacy. These include directions to prevent deplatforming, outing, context collapse, and content theft. Because the risks of digital intimacy are typically more pronounced, in volume and severity, for sex workers\footnote{Sex work is broadly the exchange of sexual services for money~\cite{smith2018revolting}, and includes work such as escorting, stripping, webcamming (performing online private or public shows ranging from dancing to pornographic performances~\cite{jones2020camming}), sexting, and content production and sale, which includes both digital goods---e.g., images and videos---and physical goods such as used clothing. While many readers---particularly those in the United States---may associate the term sex work with illegality, we note that sex work encompasses a wide range of services, many of which are legal in a variety of jurisdictions across the world. See the Background section for further discussion.}
, we focus on digital sex work as a lens through which to understand digital intimacy threats and to understand potential research directions to mitigate them.  
By centering the most marginalized users when building solutions, we can increase safety for all.

In the remainder of this piece, we provide background on digital sex work and present \textit{threat models} characterizing the risks sex workers face. To develop these threat models, we draw on the rich academic and community-led work investigating sex workers' safety (e.g., \cite{sanders2018internet, barwulor2021disadvantaged, barakat2021community, mcdonald2021s,bhalerao2022ethics,blunt2020erased}).\footnote{The literature we draw upon is broader than we can cite; for further reading, we compiled a living reading list here: \url{https://github.com/VaughnHamilton/SW_Research}.}
We then distill a set of directions for addressing the risks of digital intimacy. We conclude by illustrating the ways in which these directions can not only benefit sex workers, but also those engaged in recreational digital intimacy
and other marginalized groups not engaged in digital intimacy but who nonetheless face overlapping risks. For example, activists and racial and gender minorities also experience deplatforming and censorship by online platforms~\cite{blunt2020posting} and people in LGBTQIA communities may desire to separate their digital personas to avoid their sexuality or gender identity being known to those who may endanger them. 

\section{Background: Sex Work and Technology}\label{sec:taxonomy}
Our goal in this section is to equip readers unfamiliar with sex work---or readers who have only encountered highly stylized and often stereotypical depictions of sex work in media---with the necessary background to understand our proposed research agenda.\footnote{We note that a deeper understanding of sex work is necessary for actively pursuing some of the proposed research; interested readers can find additional resources at \url{https://github.com/VaughnHamilton/SW_Research}.} Sex work is the exchange of erotic labor or sexual services, including production of intimate content, for money~\cite{unaids}. Despite the WHO, UN, and other human-rights bodies supporting the decriminalization of sex work, sex work is highly stigmatized and over-policed, even in countries where sex work is legal~\cite{carceral_fosta,halperin2017war}. While we do not attempt a comprehensive overview of all forms of sex work, we do highlight common ways sex workers leverage technology. We conclude by articulating how this background generalizes to those engaged in recreational digital intimacy. We include mock-ups of the described platforms in \cref{fig:sw:platforms} to help visualize these platforms.

\subsection{Sex Workers' Technology Use}

Sex workers use technology to advertise, communicate with clients, vet clients, 
offer digital services~\cite{hamilton2022risk}, find peer support and learn harm-reducing information, e.g., about health and digital security~\cite{barakat2021community, mcdonald2021s}. 

\ifdefined\hidefigures
\else
\begin{figure*}[h!t]
     \centering
     \begin{subfigure}[!t]{0.48\textwidth}
         \centering
                 \includegraphics[width=\textwidth]{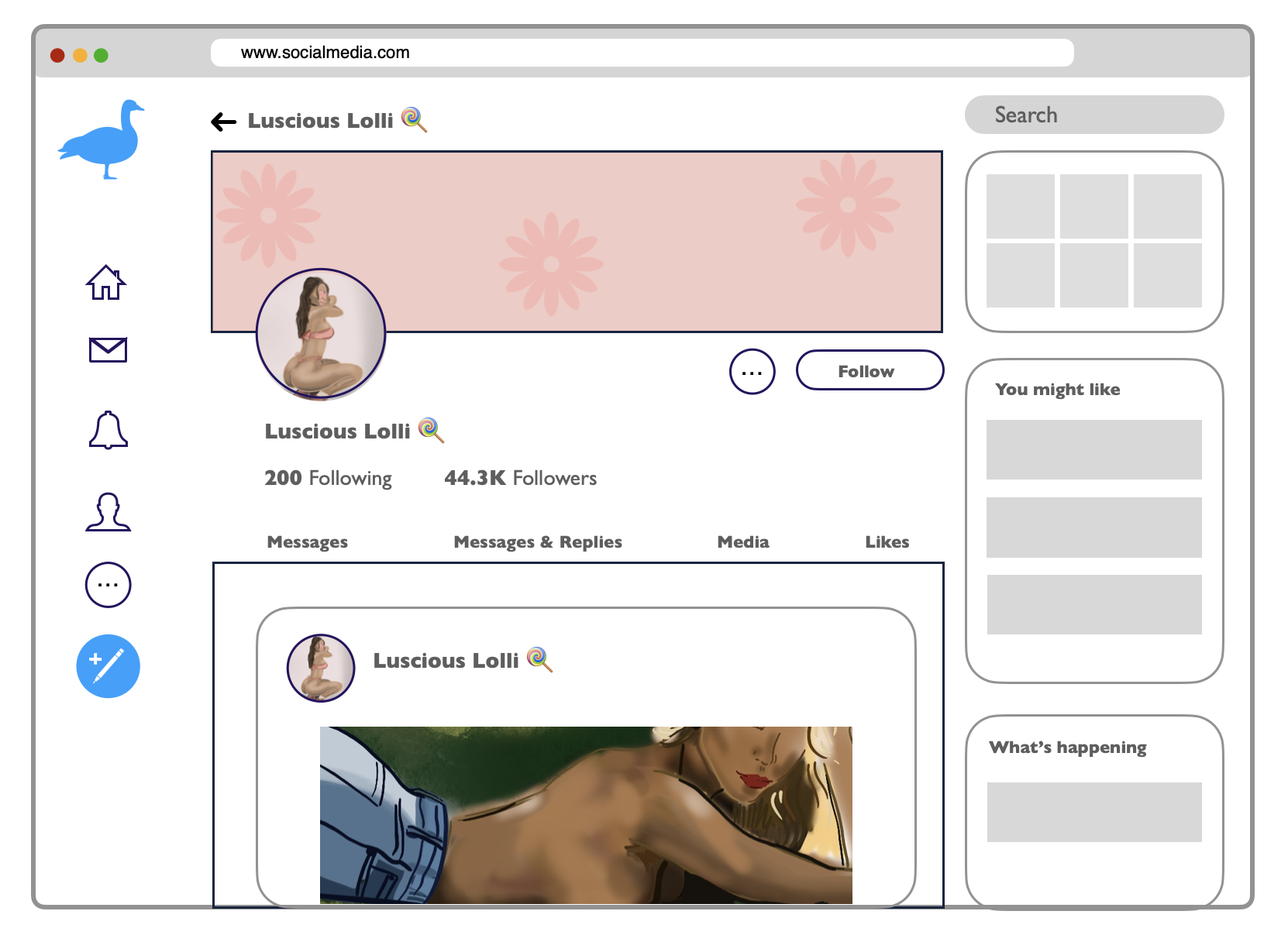}
        \caption{Social media page.}\label{fig:sw:platforms:socialmedia}

     \end{subfigure}
     \hfill
          \begin{subfigure}[!t]{0.48\textwidth}
         \centering
                 \includegraphics[width=\textwidth]{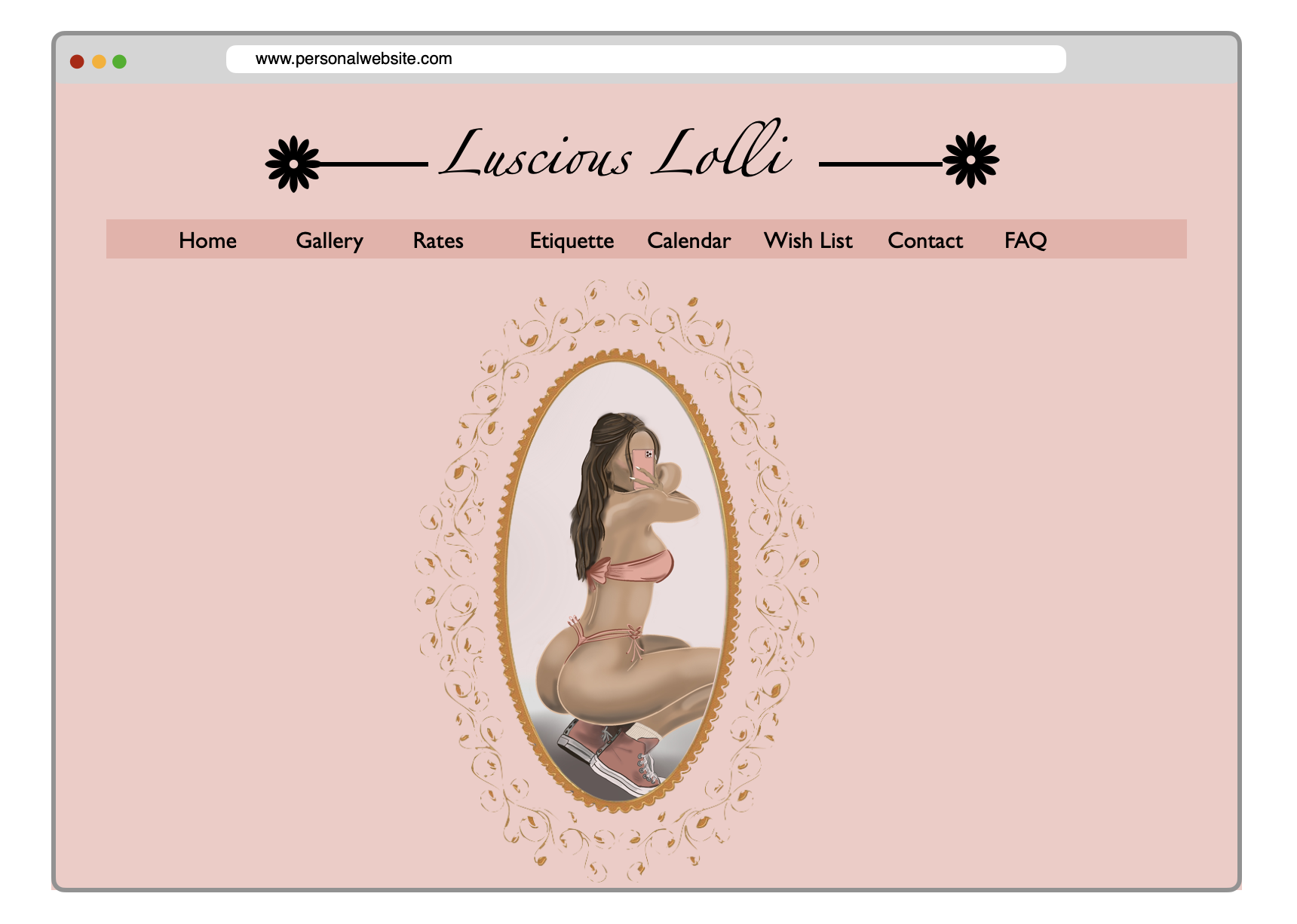}
        \caption{Personal website.}\label{fig:sw:platforms:personalwebsite}

     \end{subfigure}
     \hfill
     \begin{subfigure}[!t]{0.49\textwidth}
         \centering
        \includegraphics[width=\textwidth]{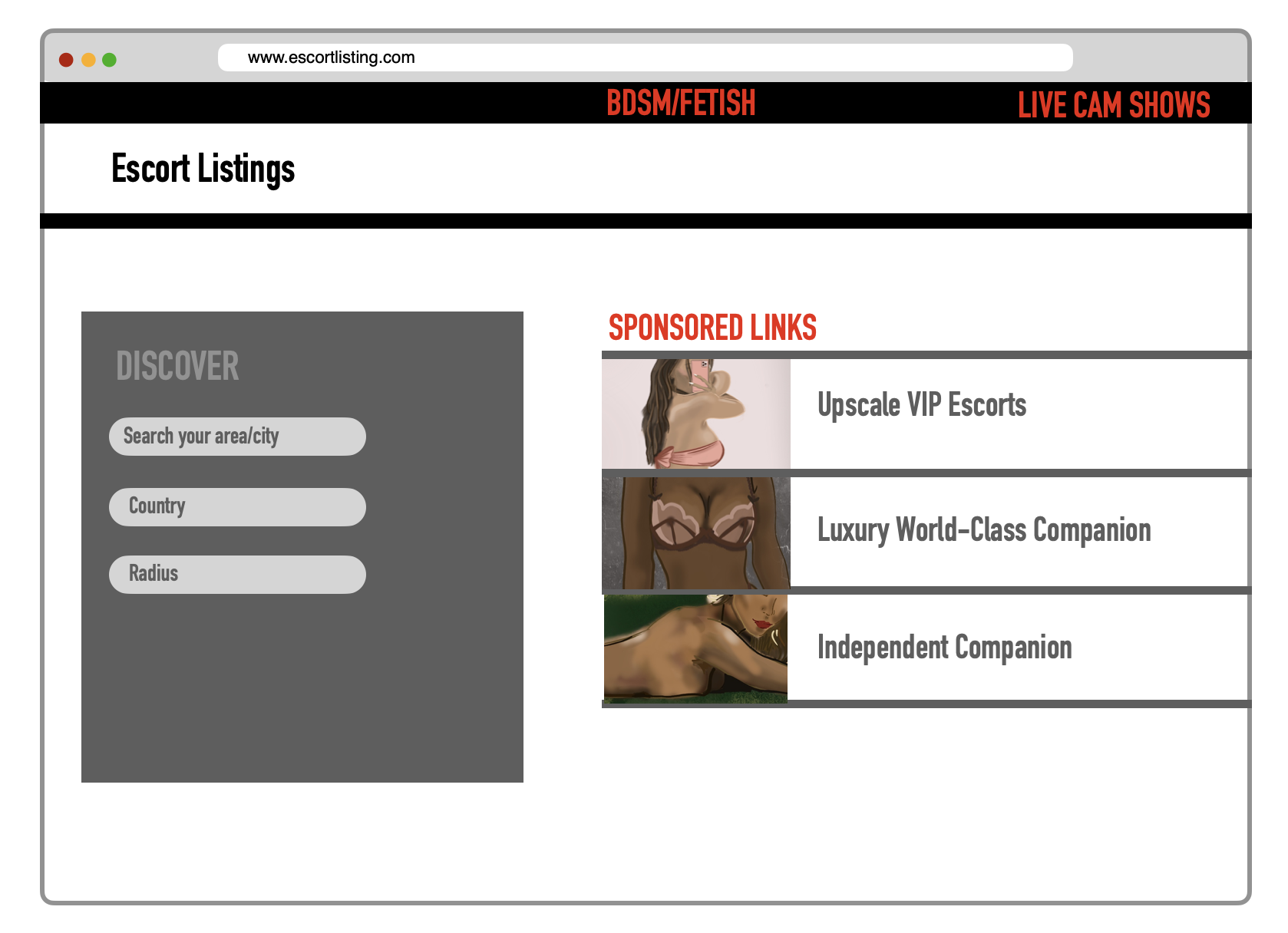}
        \caption{Escort directory.}\label{fig:sw:platforms:escort}
     \end{subfigure}
          \hfill
     \begin{subfigure}[!t]{0.49\textwidth}
         \centering
        \includegraphics[width=\textwidth]{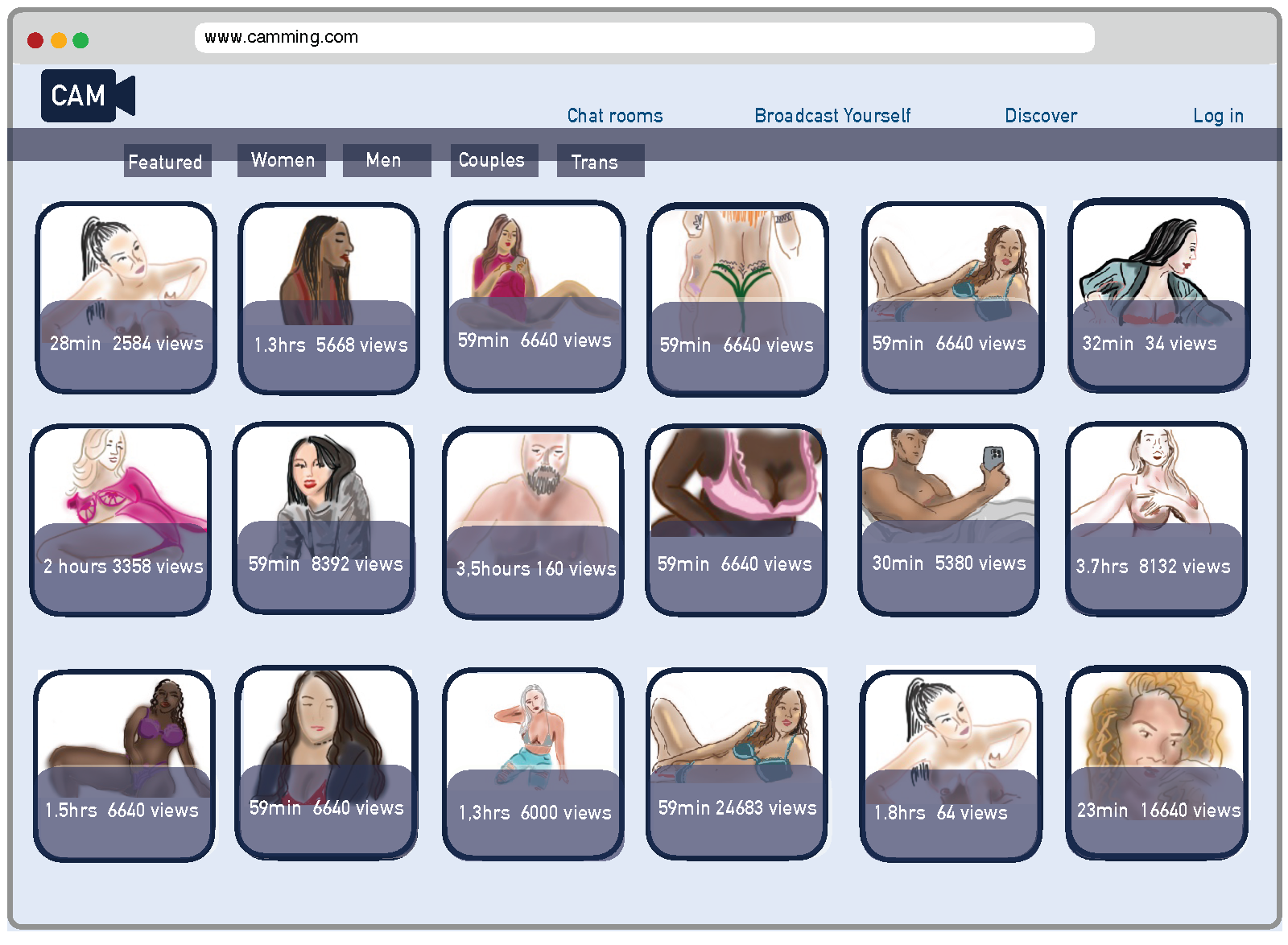}
        \caption{Camming platform.}\label{fig:sw:platforms:camming}
     \end{subfigure}
      \hfill
          \begin{subfigure}[!t]{0.49\textwidth}
         \centering
        \includegraphics[width=\textwidth]{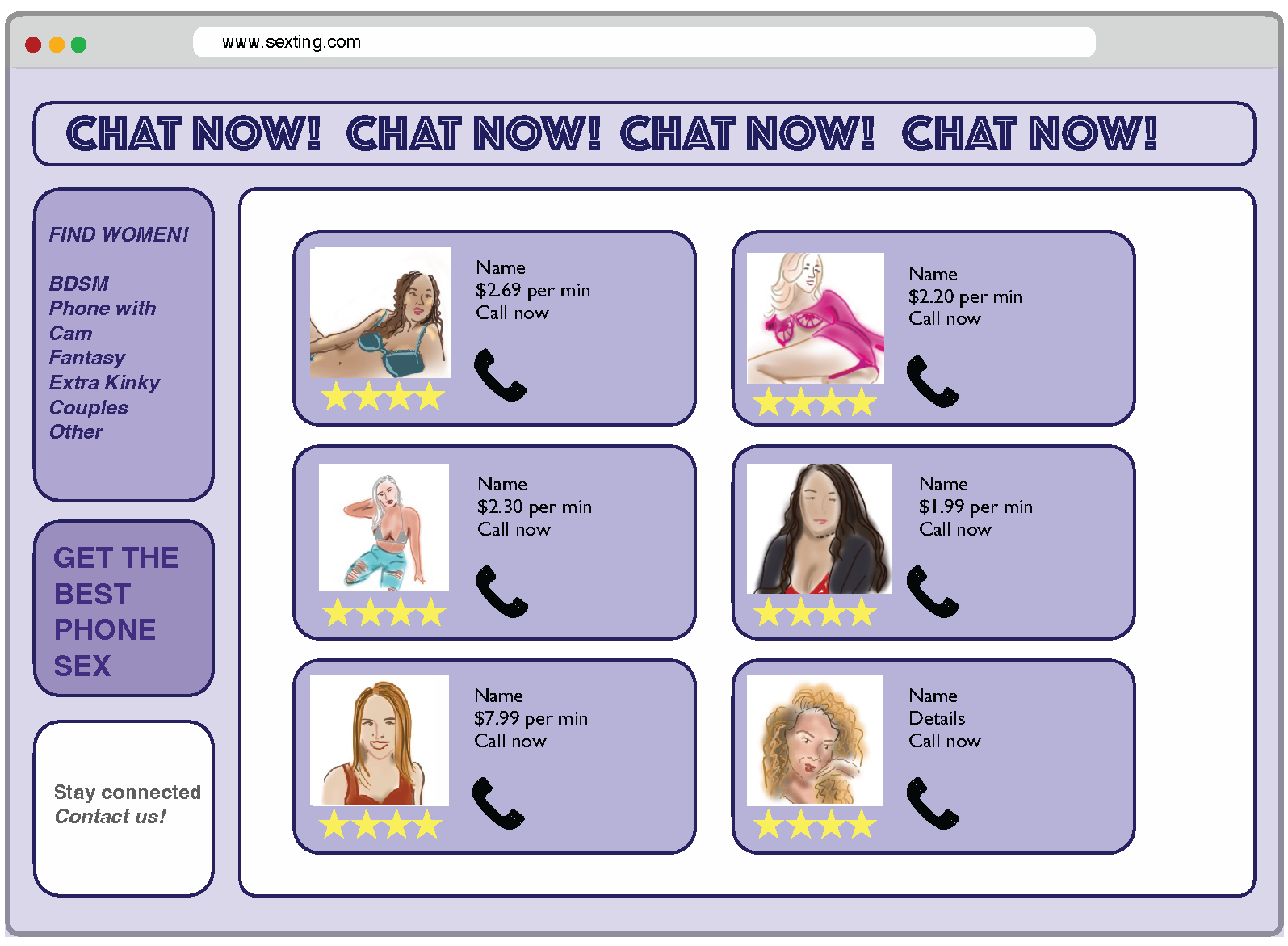}
        \caption{Phone sex website.}\label{fig:sw:platforms:phonesex}
     \end{subfigure}
     \caption{Stylized visualizations of sex work platforms.\\
     }
         \label{fig:sw:platforms}
     \clearpage
 \end{figure*}
\fi

\paragraph{Client Acquisition: Advertising.} Digital advertising is an important tool for sex workers to attract new clients (see~\cref{fig:sw:platforms}). Advertisements often contain photos and may list or allude to services, prices, and client contact procedures. Sex workers post advertisements to both mainstream and sex-work specific platforms (\cref{fig:sw:platforms:socialmedia}) and sometimes maintain personal websites (\cref{fig:sw:platforms:personalwebsite}).

Sex workers working for an employer (such as a brothel, strip club, or escort service) may leverage the organization itself to advertise (\cref{fig:sw:platforms:escort}). Alternatively, some sex workers work on digital platforms like camming sites or phone sex platforms, which may promote sex workers' content or profiles to prospective clients through advertisements on porn websites or listings directly on the camming site (\cref{fig:sw:platforms:camming} and \cref{fig:sw:platforms:phonesex}). Workers may have the option to pay to promote their listing (e.g., per click or for a particular rank in the listings). 

\paragraph{Client Acquisition: Vetting.} When sex worker's services include in-person interaction, they will generally attempt to ascertain the risk associated with meeting a new potential client before meeting in person. Vetting strategies vary but may include requiring client to provide state-issued identification, a workplace profile, or references from colleagues, either provided directly or through a vetting platform~\cite{sanders2018internet, mcdonald2021s}.  

\paragraph{Client Services.} 
When providing services in-person, sex workers leverage technology to maintain physical safety. 
Many use a practice called ``covering,'' where they share their booking information with a colleague or trusted contact---e.g., with location sharing through their phone or describing their booking details via a messaging app---and check in after the appointment. Some apps and wearables offer similar functionality but are typically aimed at recreational online dating.\footnote{See, for example, Flare: \url{https://getflare.com/}.} 

Technology is also involved in creating and distributing intimate digital media (e.g., cameras, media editing applications, file hosting services, sex-work advertising platforms). For synchronous services, creation of content happens simultaneously with distribution, often mediated by a sex work platform that either offers its own video or phone calling infrastructure or links worker profiles to other video streaming platforms. 
Asynchronous services usually involve a sex worker creating video, audio, or image content (and potentially physical goods such as used clothing) and then later listing that content for sale on an adult or mainstream platform (e.g.,~\cref{fig:sw:platforms:socialmedia}) or providing access through a subscription service. Distribution of this content might be managed through the platform or manually by the sex worker (e.g., sharing password protected files hosted on cloud storage accounts). 

\paragraph{Client Communication and Client Maintenance.} 
Sex workers use many mainstream communication platforms (e.g., email, SMS, and direct messaging) to communicate with clients.  Due to the marginalized nature of sex work, some sex workers opt to use adult-specific platforms to communicate with clients or use technologies that provide stronger privacy protections, like end-to-end encrypted email services and messaging apps. Sex workers may choose privacy preserving systems because mainstream services might forbid ``illicit or immoral'' communication, regardless of whether that communication is work-related or personal. Sex workers may also use technology to communicate with clients while providing services, for example using chatbots, messaging scheduling, or even hiring human assistants or moderators to reply to messages on their behalf. 

\paragraph{Payments.} Sex workers use multiple methods for payment including cash, gift cards (digital or physical), peer-to-peer payment or banking apps, 
checks or direct deposits (e.g., via a sex work platform), and to a limited extent
direct payment processing.
Many major payment platforms forbid payments associated with sex work, regardless of whether that work is conducted legally. As a result, many transactions for in-person work are conducted in cash. 
Sex workers may also encourage clients to buy them gifts, accept gold, 
cryptocurrencies, or access to the client's own credit cards, 
bypassing interactions with payment processors or platform intermediaries altogether.  Workers' choice of payment method is influenced by client preferences, workers' need for privacy, and what payment apps or platforms the worker can access.

\paragraph{Community.} Sex workers also use technologies such as groups in messaging apps and mainstream and adult forum platforms to communicate with each other. Their primarily text-based communications range from water-cooler workplace conversation, to peer support for serious safety issues, to organizing for advocacy and justice~\cite{barakat2021community}. People engaged in recreational intimacy also obtain peer support for issues around dating and also engage in sexual expression as part of online communities. 

\subsection{Recreational Intimacy}
People engaged in recreational digital intimacy use technology in many of the same ways. Those who are online dating or seeking partners to sext with engage in similar vetting practices to those described above in Client Acquisition, leveraging connections such as friends or online groups in the absence of colleagues. The same covering behavior as described in Client Services may be seen among online daters; those creating and sharing intimate content recreationally may do so either in 1:1 situations or one-to-many situations, thus using many of the same technologies mentioned above for creating, distributing, and attempting to maintain control over their intimate content. Finally, people engaged in recreational intimacy use many of the same platforms as described in Client Communications to share intimate content, and may chose to use privacy-preserving communication systems for the same reasons.

\section{Mechanisms of Harm}
\label{sec:surfaces}

Safety encompasses multiple interdependent dimensions, including physical safety, financial security, privacy, and access to community. Technology can both cause and facilitate violations to these safety needs. We identify four broad \textit{mechanisms of harm}:
\begin{itemize}
    \item[--] \textbf{Deplatforming}: having an account or content removed or suppressed (e.g., shadowbanning~\cite{blunt2020posting}) from a digital platform. This can threaten financial security if a worker loses their advertising base and cannot conduct business 
    or access funds 
    and prevents access to community if a worker is not able to connect with others.

    \item[--] \textbf{Payment Inoperability}: inability to receive payment from clients or pay for necessary goods and services because payment formats are incompatible
    . This can prevent workers from being able to pay for essentials such as food, housing, and medical care (for example, if their primary income is in the form of gift cards), and may lead to privacy violations by forcing a sex worker to use a payment form that reveals personal information (e.g., bank transfer).

    \item[--] \textbf{Outing \& Context Collapse}: having one's intimate content, identity as a sex worker (or related information such as sexual identity), or personal information (e.g., address, legal name) exposed without consent. Such privacy violations can threaten emotional well-being as well as physical safety in cases of stalking or being outed to family.

    \item[--] \textbf{Content Theft}: having content (e.g., images, videos, or ad copy) stolen and republished without consent
    .\footnote{A person's initial act of sharing intimate content, either as work or recreation, does not condone later unfettered sharing on the part of the initial recipient.  Similarly, engaging in intimate activity in one context (i.e., on a sex work platform) does not automatically indicate consent to have that information made public.} This can harm a worker's emotional well-being and in turn threaten physical safety, for example if a client is led to expect a service the worker does not offer from a fraudulent ad. Content theft also affects financial safety by directly stealing revenue from the worker, and can violate privacy if images are republished in a place that is likely to out the worker.
    \end{itemize}

\subsection{Case Studies}
The following stylized case studies of sex worker experiences 
illustrate these mechanisms of harm. These case studies were carefully constructed by our community consultant from academic literature\footnote{\url{https://github.com/VaughnHamilton/SW_Research}} and real sex worker experiences, while avoiding real details from any one source to minimize risk.

\medskip
\noindent\textit{Case Study 1:} 
K spent the last few summers as a food server. They were surprised to find food service wasn't quite what they expected. They got tipped better when they flirted with customers and dressed in ways that aligned with gender- and beauty-norms.  
 On <social media platform>, K saw people talking about selling nude photos through <social media platform>. 
They felt selling photos was not much different from their food service work and might pay better, so they decided to try it. They created a new account and successfully started selling nudes directly via DM. After a few months, however, they were \harmBox{deplatformed}: their account was deleted even though they had not violated the terms of service. K lost their source of income and contact with many of their followers. They had heard from a few friends about <sex work platform> where accounts were more stable and people could more easily buy their content and even buy a subscription. K decided to start an account on this platform and became very successful.

K is comfortable with what they do and has made quite a lot of money doing it. But, K worries about their lack of control over their content. To avoid \harmBox{content theft}, K watermarks their content and sets up a Google alert for their stage name to see if their content has been reposted. They also try to avoid \harmBox{outing \& context collapse} by keeping people in their personal networks from finding their account: they proactively block personal contacts on <sex work platform>. Eventually contacts did find their content and posted on another site to harass them. They ignored the haters, but wish there was some way they could protect themselves more thoroughly.

\medskip
\noindent \textit{Case Study 2:} 
M has provided BDSM services to clients in-person for many years. They are a permanent resident of the country in which they work and legally registered to run their BDSM business. They immigrated from a country where being queer is criminalized. If anyone from their home country were to learn about their business and \harmBox{out} them, it might 
create a risk of violence for their family still living there and prevent M from returning. 
Thus, they take great care---using multiple different phones, wiping any relevant information from their phone before crossing borders, being vigilant about their social media posts and likes---to ensure nothing connects their personal identity to their work.
Despite their careful management, M has experienced \harmBox{content theft}: the photos from their ads have been copied and reposted on other sites. They can't find a way to get them taken down and M is afraid of who might see them. 
M wishes they could be ``out'' about their sex work within the country in which they live so they didn't have to stress so much about their photos
and participate in activism to improve working conditions for other sex workers, but the risk is simply too high. 

\medskip
\noindent \textit{Case Study 3:} 
R is a single parent and has custody of her two toddlers every other week. She needs a source of income that allows her to spend that whole week with her children. In the week when she doesn't do childcare, she has experimented with different kinds of sex work to try to find the one with the most flexibility and highest income. She crossed town borders to strip outside her community, worrying that if someone who knew her found out they might \harmBox{out} her to her ex-husband and she could lose custody. Escorting is legal in R's country. She sometimes accepts high value jobs from a discreet escort agency that doesn't require her to share pictures of her face. When these jobs were slow, she opened a content sale account, sharing heavily-edited lewds and nudes of herself, but she blocks her whole home country and the country where her ex lives. She promotes her content account lightly on social media 
and carefully removes metadata from all her photos. 
She still has problems with \harmBox{payment inoperability}: the <peer-to-peer payment app> she uses to sell her content has disabled her account multiple times, and more than once she has almost missed rent because she couldn't withdraw the money she had already earned.

\subsection{Harm Surfaces}
Prior work (e.g., \cite{sanders2018internet, barwulor2021disadvantaged, barakat2021community, mcdonald2021s,bhalerao2022ethics,blunt2020erased}) reveals a number of ways in which the technologies sex workers use can be a source of harm.
Here, we summarize the categories of \emph{technological interfaces} involved in sex work that require implicit trust relationships (illustrated in \cref{tab:taxonomy}).

     \ifdefined\hidefigures
\else

\begin{table*}[h!t]
 \caption{A summary of the technologies used by sex workers and the purposes for which they are used, as well as the mechanisms through which they may cause harm. 
 }
\resizebox{\textwidth}{!}{
  \centering
    \begin{tabular}{l|l|l|l}
      \toprule
& \multicolumn{1}{|c}{\textbf{Business Technologies}} &  \multicolumn{1}{|c}{\textbf{Safety Strategies \& Technologies}} & \multicolumn{1}{|c}{\textbf{Mechanisms of Harm}}
\\

      \midrule

            \multicolumn{1}{c|}{\rotatebox[origin=c]{90}{\textbf{Client Acquisition}}} 
            & 
            \multicolumn{1}{c|}{\makecell[l]{Personal Website\\Mainstream Platforms\\~~Social Media\\Sex Work Platforms\\Devices\\Media}} 
           &
           \multicolumn{1}{c|}{\makecell[l]{Vetting via\ldots\\~~Professional Credentials (e.g., work ID)\\~~References (e.g., from other sex workers)\\~~Sex Work Platforms or Bad Client Lists\\Paywalls\\Age Gates\\Using Multiple Accounts \& Devices\\Self-Censorship}}
            &
            \multicolumn{1}{c}{\makecell[l]{Deplatforming \\ Content Theft \\ Outing \& Context Collapse}}\\  
            
         \\

      \midrule

       \multicolumn{1}{c}{\rotatebox[origin=c]{90}{\textbf{Client Services}}} 
       & 
   \multicolumn{1}{|c}{\makecell[l]{Payment Platforms\\Mainstream Platforms\\~~File Sharing Services\\~~Streaming Services \\Sex Work Platforms\\ Devices \\ Media}}
   &
   \multicolumn{1}{|c}{\makecell[l]{Covering via\ldots\\~~Alarm Apps\\~~Messaging Apps \& SMS\\Security Cameras \\On-Platform Blocking\\Preventing Unauthorized Sharing via...\\~~Google Alerts\\~~Incorporating Client Name into Media\\~~Watermarking\\Paywall\\Age Gate}}
   &  
   \multicolumn{1}{|c}{\makecell[l]{Deplatforming \\ Payment Inoperability\\ Content Theft \\ Outing \& Context Collapse}} \\

       \midrule

           \multicolumn{1}{c}{\rotatebox[origin=c]{90}{\textbf{Client Communication}}} 
       & 
   \multicolumn{1}{|c}{\makecell[l]{Mainstream Platforms\\~~Email\\~~SMS \& Messaging Apps\\~~On-Platform Direct Message\\~~Social Media\\Sex Work Platforms\\~~On-Platform Direct Message\\~~Live chat (e.g., during a virtual show)\\~~Chatbots\\Devices\\Media}}
   &
   \multicolumn{1}{|c}{\makecell[l]{Using Multiple Accounts \& Devices\\Encryption (Email \& Messaging Apps)\\Limit Available Communication Channels\\Blocking\\Human or Automated Moderators}}
   &  
   \multicolumn{1}{|c}{\makecell[l]{Deplatforming \\ Content Theft}}     
           \\

   \midrule
     \multicolumn{1}{c}{\rotatebox[origin=c]{90}{\textbf{Payments}}} 
       & 
   \multicolumn{1}{|c}{\makecell[l]{Cash\\Gifts/Gift Cards (Wishlists)\\Cryptocurrencies\\Payment Platforms\\~~(e.g., Venmo, Cashapp)}}
   &
   \multicolumn{1}{|c}{\makecell[l]{Using Multiple Accounts \& Devices\\Self-Censorship}}
   &  
   \multicolumn{1}{|c}{\makecell[l]{Deplatforming \\ Payment Inoperability \\ Outing \& Context Collapse}} 
        \\

    \midrule
     \multicolumn{1}{c}{\rotatebox[origin=c]{90}{\textbf{Community}}} 
       & 
   \multicolumn{1}{|c}{\makecell[l]{Mainstream Platforms\\~~Messaging Apps\\~~Social Media\\Sex Work Platforms\\~~Forums}}
   &
   \multicolumn{1}{|c}{\makecell[l]{Using Multiple Accounts \& Devices\\Self-Censorship}}
   &  
   \multicolumn{1}{|c}{\makecell[l]{Deplatforming}} 
        \\

      \bottomrule
    \end{tabular}
  }
  \label{tab:taxonomy}
\end{table*}
\fi

\paragraph{Devices.} 
Sex workers have digital interactions with a wide number of people, some of whom are interested in stalking, doxxing, or obtaining information from them. As a result, their devices (e.g., cell phones, computers, and/or cameras) may become infected with malware, spyware, or ransomware. Device compromise may out a worker by exposing personal information or leaking intimate media.

In addition to device compromise, those working in-person may be harmed by their clients' devices, which could be used to secretly record their engagement and distribute the resulting content without the worker's knowledge or consent. 
Further, a sex worker may be forced to surrender devices (e.g., when crossing international borders),
which may out them as a sex worker.

Finally, devices may pose usability challenges. For example, to avoid potential cross over of identifying information (IP address, MAC address, Bluetooth identity, etc.) between sex work and non-sex work accounts and profiles some sex workers use multiple devices~\cite{mcdonald2021s}, a burdensome practice that is difficult to implement perfectly.

\paragraph{Media.} 
Media (images, videos, audio recordings, personal webpages, advertising copy) can be a source of harm when they are non-consensually produced or shared.  
Further, contextual information (e.g., image backgrounds) and metadata may be used to out a worker.

\paragraph{Mainstream Platforms.} 
Many mainstream platforms have a history of selectively enforcing policies to the detriment of sex workers~\cite{barwulor2021disadvantaged, hamilton2022risk}. This includes abruptly deplatforming sex workers  
or failing to support sex workers targeted by online harassment and abuse. 
Because platforms rarely provide clear guidance on what content will lead to deplatforming, sex workers must gamble on what content they can share on mainstream platforms to maximize their followings without risking deplatforming.

Mainstream social media platforms also pose privacy risks for sex workers. A worker may be outed by a platform recommending they connect with clients on their personal social media, or advertising their work accounts to family. While there are strategies for preventing this (e.g., proactive blocking, as K did in Case Study 1), they are not 100\% effective. Context collapse is difficult to predict and prevent because there is very little transparency around the data aggregation and prediction tools used by these platforms.

\paragraph{Sex Work Platforms.} 
Sex work platforms may fail to implement protections to prevent content theft or to allow workers to effectively screen and remove harassing clients. 
Additionally, platforms often require government ID and/or legal name and address from workers for age verification and payment. Breaches of platform information could out a worker by leaking these details, particularly linked together with a worker's work persona and/or content, which are easily identifiable to their clients and may also out them to others in their life.\footnote{An example of a data breach of sex worker information was Pornwikileaks, which no longer exists \url{https://www.cnet.com/culture/pornwikileaks-reveals-identities-of-porn-stars/}}

\smallskip \noindent
\emph{Scraping.} The threat of scraping, a consequence of aggregating sex-work-related content, is significant for sex workers. There is evidence that sex-work-related content is regularly compiled in massive databases. For example, anti-trafficking organizations, corporate entities, or law enforcement may scrape workers' personal websites and other content to create databases of worker information for use in criminal prosecution, mass-text campaigns, deplatforming,\footnote{For example, see this patent: \url{https://patents.google.com/patent/US10019653B2}.} and border control~\cite{bhalerao2022ethics}.

\paragraph{Payment Platforms. }
Payment apps may become inoperable, freeze payments, or deplatform sex workers. This is true regardless of the legality of the services paid for and is not required by U.S.\ law.\footnote{See analyses of existing legislation and the impacts of payment deplatforming such as \url{https://www.nswp.org/sites/default/files/fosta_briefing_note_2018.pdf}, \url{https://www.aclu.org/news/lgbtq-rights/how-mastercards-new-policy-violates-sex-workers-rights}, \url{https://oneill.law.georgetown.edu/unpacking-the-dangers-of-mastercards-push-to-exclude-sex-workers-from-safer-sex-trade-spaces/}.} While it is not known how precisely payment processors flag payments, hypotheses include keywords in the notes provided with payments, patterns of payments, and networks of payment connections. In the U.S., specifically, banks may also freeze or delay sex workers' payments or may close their accounts (often retaining the money stored in them) altogether.

\section{Concrete Opportunities for Technical Research}%
\ifdefined\hidefigures
\else
\begin{table*}[]
\begin{tabular}{p{3cm}p{3cm}p{11cm}}
\textbf{Mechanism of Harm}                         & \textbf{Mitigation Direction}              & \textbf{Exemplar Related Work}                                                                                                                                                                                        \\ 
\midrule
\multirow{4}{*}{Deplatforming} & Filter Analysis                            & \textit{Model Extraction:} Tramer et al. 2016, Wang \& Gong 2018, Yu et al. 2020, Xun et al. 2022                                                                                                                                                                 \\
\\
 & Facial Recognition Circumvention           & \textit{Existing Techniques:} Ramachandra \& Busch 2017, Chatzikyriakidis et al. 2019, Shan et al. 2020, Radiya-Dixit et al. 2021, Wenger et al. 2021, Kelly et al. 2022, Wenger et al. 2023  \\ 
 \midrule    
\multirow{3}{*}{\shortstack[c]{Payment\\Inoperability}} & Usable Privacy-Preserving Cryptocurrencies & \textit{Documenting Lack of Usability:} Ramadhan \& Iqbal 2018, Mai et al. 2020, Moniruzzaman et al. 2020, Albayati et al. 2021, Fr\"{o}hlich et al. 2021, Jang et al. 2021, Nadeem et al. 2021, Voskobojnikov et al. 2021, Fr\"{o}hlich 2022, Ghesmati et al. 2022  \\
\midrule
\multirow{13}{*}{\shortstack[c]{Outing \&\\Context Collapse}}       & Automated Blocking of Contacts             & \textit{Documenting Privacy Risks in Social Networks:} Gross \& Acquisti 2005, Becker \& Chen 2009, Liu \& Terzi 2010, Ramachandran et al. 2012, Aghasian et al. 2017, Pensa et al. 2019  \\  \\  
                                                   & Robust Image Modification                  & \textit{Documenting Insufficiency of Existing Techniques:} Hill et al. 2016, McPherson et al. 2016, Vishwamitra et al. 2017, Niu et al. 2021. \textit{Image Deidentification:}
                                                   Fan 2018, Fan 2019, Li \& Lin 2019, Chen et al. 2021  \\ \\
                                                   & Privacy-Preserving Multi-Profile Support   & \textit{Software Multi-Profile Support:} Chen et al. 2011, Android Documentation. \textit{Documenting Insufficiency of Existing Techniques:} Habib et al. 2018, McDonald et al. 2021. \textit{Hardware Multi-Profile Support:} Armando et al. 2014, Kanonov \& Wool 2016                                                                                                                                                                              \\ \\
                                                   & Verification of Privacy Settings           & \textit{Documenting Problems with Privacy Settings:}  Yamada et al. 2012,  Humbert et al. 2013, Yu et al. 2016,  \textit{ Verifing Privacy Policies:}
                                                   Li et al. 2006, Slavin et al. 2016, Andow et al. 2020
                                                    
                                                   \\
                                                   \midrule
\multirow{8}{*}{Content Theft}                                      & Certificate Infrastructure for Content     & \textit{Digital Rights Management:} Becker et al. 2003, Liu et al. 2003, Subramanya \& Yi 2006                                                                                                   \\                                                       \\
                                                   & Robust Content Matching                    & \textit{Perceptual Hashing:} David \& Rosen 2018 (Blog), Farid 2021, Apple CSAM Detection Technical Summary 2021.                                                                                         \\ \\
                                                   & Anti-Theft Technology                      & \textit{Existing Techniques:} Encrypted Media Extensions, W3C 2019. \textit{Cryptographic Watermarking:} Dittmann 2001, Mohan \& Kumar 2008, Wan et al. 2022                                                                                                                       
\end{tabular}
\caption{A summary of exemplar prior work related to the research directions we review \& synethsize.}
\label{tab:related}
\end{table*}
\fi

We review \& synthesize several opportunities for research to support safer digital intimacy. We encourage researchers to carefully consider the regulatory and legal landscape in which they work before pursuing the directions synthesized in this article; not all directions may be appropriate for all contexts and researchers should take care in determining which populations to target their interventions toward (e.g., those doing particular forms of intimate work, those engaged in recreational digital intimacy, or other groups). In Table~\ref{tab:related} we point to related literature, as a starting point for embarking on research in each direction. 

\subsection{Deplatforming}

As explained above, deplatforming is a significant risk to sex workers, 
which suggests multiple research directions, including:

\ifdefined\maglayout
\smallskip \noindent \textbf{(1) Filter Analysis.}
\else
\begin{itemize}
  \item[(1)] \textbf{Filter Analysis.}
\fi
Payment platforms and other mainstream platforms deploy automated scanning technology that flags accounts it determines are associated with sex workers, regardless of whether the flagged users' work is legal or the flagged users are using the platforms strictly for personal use. The same technologies is easily adapted to identify other groups deemed undesirable or troublesome, including activists and those with disabilities.\footnote{For example, see this patent filed by AirBnB, which seeks to identify sex workers and those with disabilities: \url{https://patents.google.com/patent/US9070088B1}.}  
Unfortunately, existing uses of filtering technologies are not well understood. An intriguing line of research would be generating tools that automatically test the behavior of a filter and continue to adapt to the filter as it changes. 
  Platforms themselves could even consider offering filter pre-screening tools, as content creators and platforms may be aligned in their goals: content-creators seek not to lose their accounts over disallowed content and platforms seek to avoid having such content on the platform~\cite{hamilton2022risk}. However, for tools designed to prevent algorithmic profiling, creating and maintaining such a tool once platforms become aware of its existence will be challenging.

While there exists significant work on filter analysis---and model extraction in particular---in the academic literature, \emph{tools} that can automatically learn filter behavior are lacking. There are several technical challenges in constructing such tools, including API rate limiting and avoiding detection of adversarially crafted queries.  Moreover, many content moderation pipelines involve human intervention after content has been automatically flagged, which would result in arbitrary moderation decisions for content near the edge of a moderation decision boundary.

  \ifdefined\maglayout

    \smallskip \noindent
    \textbf{(2) Facial Recognition Circumvention.}
  \else
    \item[(2)] \textbf{Facial Recognition Circumvention.}
  \fi
  In addition to traditional filters, facial recognition technologies are used to automatically identify and filter sex workers and other ``undesirable'' persons from using digital platforms. Research in several directions has thus begun on ``anti-facial recognition''. These include approaches to prevent scraping of images for unauthorized use in training facial recognition algorithms or assembling databases of ``undesirable'' people. There are several limitations to the existing state of the art, including limits on usability. In some cases, techniques make unreasonable assumptions, eg. access to the model or the ability to modify the training set, which may be impossible in practice.  Other academic approaches require modifying the \emph{appearance} of images, e.g., using disguises, for de-identification or developing techniques for de-identifying static images; noticeable image modification may be incompatible with sex worker's business interests.  Image modification may also be unreasonable for other users, eg. activists who need to be recognizable as part of their organizing.  

  Further research is needed to develop usable adversarial machine learning techniques for image and video content that help users protect themselves from facial recognition algorithms.\footnote{Efforts have already begun in both directions. Patents for such malicious uses of facial recognition have been field, such as \url{https://patents.google.com/patent/US10019653B2} and research efforts have been funded to combat such uses, e.g., \url{https://www.nsf.gov/awardsearch/showAward?AWD_ID=2144988}.}  Importantly, any resulting tools need to be very unobtrusive in order to not undermine the value of the content.

 \ifdefined\maglayout
 \else
\end{itemize}
\fi

\subsection{Payment Inoperability}
Frozen or delayed payments endanger sex workers' livelihoods. Sex workers are already highly innovative in discovering payment methodologies that are not subject to tracking or delays. However, as clients increasingly move to traceable digital payment technologies, there is a need for further research. For example, we suggest: 

  \ifdefined\maglayout

    \smallskip \noindent
    \textbf{(3) Usable Privacy-Preserving Cryptocurrencies.}
  \else
  \begin{itemize}
    \item[(3)] \textbf{Usable Privacy-Preserving Cryptocurrencies.}
  \fi
   Non-traditional, distributed, digital payment platforms, such as cryptocurrencies, are in theory a promising tool for sex workers.  Many cryptocurrencies offer either formal privacy properties or at least pseudonymity.  Despite the potential of this technology---and the significant resources the security and privacy community has devoted to developing cryptocurrencies---cryptocurrencies are rarely used by sex workers. There are significant usability barriers for these privacy-preserving systems; sex workers must pay for goods and services with the payments they receive, which is challenging with cryptocurrencies.  Moreover, many clients may be less technologically sophisticated or unwilling to use cryptocurrencies.  While recent work has started to examine usability concerns, we are unaware of significant work trying to integrate these usability insights up the technology stack.  If a client is unable to understand how to purchase and send cryptocurrency, it is impractical for sex workers to demand its use. Additionally, cryptocurrency valuations are highly volatile and exchanging cryptocurrencies risks outing a sex worker.  Finally, in cases where sex workers do chose to accept cryptocurrencies, they may be deplatformed if they use online wallets designed to make cryptocurrencies usable by non-experts.

 \ifdefined\maglayout
 \else
\end{itemize}
\fi

\subsection{Outing and Context Collapse} 

Due to the stigmatized nature of sex work, many sex workers want to maintain tight control over their identity as a sex worker. The increasingly digital nature of sex work makes this difficult. Many sex workers publicly share images of themselves on both mainstream and sex work platforms. While such images might be necessary as advertisements, they significantly increase the risk that the sex worker might be outed to their personal community. This issue suggests many different research directions:

  \ifdefined\maglayout

    \smallskip \noindent
    \textbf{(4) Automated Blocking of Contacts.}
  \else
  \begin{itemize}
    \item[(4)] \textbf{Automated Blocking of Contacts.}
  \fi
 One technique sex workers use to minimize their risk of being outed is to proactively block friends and family on work-related social media accounts. However, it is not clear what the \emph{best} approach to proactive blocking would be. Should sex workers only block personal contacts? What about the social media ``friends'' of those contacts? Creating an automated blocking tool, configurable according to a sex worker's personal needs, would reduce the effort and anxiety associated with this task. Moreover, measuring the privacy utility of blocking different sets of contacts could provide better transparency into the effectiveness of this approach. 
 Prior work has aimed to measure the privacy leakage of different social network compositions and to provide tools and metrics to reduce privacy risk for individual network users. However, to our knowledge, these prior works do not address the combination of needs presented by our participants: the need for a usable, privacy preservation tool appropriate for protecting single-person multi-accounts (e.g., where a single person has two accounts in the same social network but wants those accounts not to be connected to each other), potentially across several social sites.  We also note that this problem has significant connections to graph sketching problems in theoretical computer science, as tools may not be able to easily access a view of the full social network.

   \ifdefined\maglayout

    \smallskip \noindent
    \textbf{(5) Robust Image Modification.}
  \else
    \item[(5)] \textbf{Robust Image Modification.}
  \fi
 Sex workers who hope to keep their identity secret often blur their faces and other identifying features to prevent contacts from recognizing them.  Such efforts seek to prevent \textit{human} identification, rather than \ \textit{machine} identification; the latter is addressed in direction (2).
 It would be valuable to rigorously understand the real privacy value of this approach. For example, it may be possible for a machine learning algorithm to predict the facial features (or even identity) of a sex worker, even when their content is blurred. Studying the most effective way to shield a sex worker's identity and creating tools that automatically, optimally scrub identifiable features from content could be very valuable.  Such tools would also be valuable to individuals engaging a recreational digital intimacy but hoping to mitigate the risks associated with unauthorized content sharing.

As mentioned above, there is work in modifying images such that machine learning classifiers cannot automatically detect the identity of a person in a photo at scale; importantly, in this research direction we are concerned about adversaries who are attempting to re-identify the individual in a \emph{specific} image, either through manual investigation or machine learning powered technique.  There are several works documenting that ``typical'' image modification, like blurring, may be insufficient against a motivated attacker and another line of works studies image deidentification, but it is unclear if these tools are practical to deploy or provide the requisite level of privacy required for this high-risk context.

     \ifdefined\maglayout

    \smallskip \noindent
    \textbf{(6) Privacy-Preserving Multi-Profile Support.}
  \else
    \item[(6)] \textbf{Privacy-Preserving Multi-Profile Support.}
  \fi
   Past research has focused on access control to separate user profiles on a single device or across devices. However, there is a lack of deep analysis on whether such approaches are sufficient and usable in-the-wild to protect identity in cases where a single person wants to heavily utilize two completely separate, unconnected personas.

 \begin{itemize}
            \item[(a)] \textbf{Software approaches.} Existing software approaches to keeping identities or personas separate include browser profiles and Android profiles. However, prior work suggests that such profiles may be insufficient to fully protect user identity both from social exposure and algorithmic exposure, and that marginalized users such as sex workers do not trust such software-level strategies. Future research is necessary to consolidate such approaches and increase user trust.
            \item[(b)] \textbf{Hardware approaches.} While there is existing research and deployed technology that allow for separates profiles (with different data and apps) on a given phone\footnote{\url{https://support.google.com/work/android/answer/6191949}}, these are not designed to ensure that the two separate profiles appear to e.g., browser fingerprinting services as two separate people. Further, such isolation models have not been evaluated for usability. Thus, future work is necessary to develop end-to-end approaches combining both hardware and software protections to allow users to engage as two separate ''people'' on a single device. 
            \end{itemize}

\ifdefined\maglayout

    \smallskip \noindent
    \textbf{(7) Verification of Privacy Settings.}
  \else
    \item[(7)] \textbf{Verification of Privacy Settings.}
  \fi
 Many techniques currently used by sex workers to prevent outing rely heavily on platforms' privacy protection mechanisms functioning correctly.  However, the effect of many privacy preferences is completely invisible to users and impossible to verify in practice, making it difficult for end-users trust. 
 This distrust could be overcome by designing privacy preferences that are \emph{auditable}, either directly by a user or through third-party software.  Sex workers, or trusted organizations within the sex worker community, could then verify that the protections provided by certain privacy configurations meet sex worker needs.  Importantly, the auditing mechanism must simultaneously (1) be robust, in that the auditing methods must be significantly more convincing than just a UI change, (2) not allow auditors access to previously protected information, and (3) be sufficiently understandable that it engenders trust.  We note that this idea can extend beyond privacy preferences; for example, it would be valuable to verify that metadata removal software and services actually accomplish their goal.

Prior work has aimed to verify that the functionality of e.g., business processes and Android apps aligns with their declared privacy policies. However, to our knowledge, there exists a gap in work offering trustworthy verification of the functionality of user-set controls such as privacy settings on social network platforms.

 \ifdefined\maglayout
 \else
\end{itemize}
\fi

\subsection{Content Theft}

There is a real and pressing need for techniques that give sex workers greater control over their media. Note that this is not a traditional access control problem---the capacity to manipulate or share content is unavoidably shared when workers post their content publicly or share it with a client. 
Instead, tools must enable a sex worker to \emph{detect} when their content is shared without permission or allow a sex work platform to proactively scan content to determine if it being posted with permission. Critically, these tools must be \emph{usable}, which requires a low false positive rate and built-in protections against spamming.

While several digital rights management (DRM) solutions exist, there are notable differences between this setting and the archetypal DRM setting.  Most DRM solutions are used by highly resourced organizations attempting to control a few high value pieces of content (e.g., blockbuster movies or TV shows).  These organizations have the time and expertise to enforce those rights.  The situation we consider is the opposite: there are a large number of sex workers creating a vast amount of content. Sex workers often lack the resources---in terms of time or money---to effectively assert their content rights; indeed, platforms, not fearing reprisal from marginalized sex workers, might actively make it difficult to enforce content rights.  Moreover, the nature of the content produced is quite different: while it is difficult for an individual to assert that they own a well-known film, it may be difficult to ascertain who owns the content produced by sex workers just by looking at it.  As such, applying traditional DRM solutions may not efficiently address the issue of content theft in digital sex work. Research directions include: 

\ifdefined\maglayout

    \smallskip \noindent
    \textbf{(8) Certificate Infrastructure for Content.}
  \else
  \begin{itemize}
    \item[(8)] \textbf{Certificate Infrastructure for Content.}
  \fi
 Digital sex work could be supplemented with a certificate ecosystem for media management.
        \begin{itemize}
            \item[(a)] \textbf{As infrastructure.} Sex work platforms could require that all content uploads be accompanied by documentation affirming that the content creator intended for that content to appear on the platform, such as a certificate linked to the creator that is digitally signed using standard cryptographic techniques. While such an approach would be robust, significant research is required to understand how to manage digital identity in such a system. Indeed, a thief could steal a sex worker's content and generate a false certificate claiming ownership of the content, providing the thief has control of a legitimate digital identity. Research would also need to consider whether such an ecosystem can exist while still preserving privacy for sex workers who want plausible deniability in their work.
            \item[(b)] \textbf{For proof of ownership.} Even without a strong identity management system, creating some form of certificate-based content management system could make it easier for sex workers to issue DMCA take-down requests (as the ownership of content could be verified cryptographically).
        \end{itemize}
        \ifdefined\maglayout

    \smallskip \noindent
    \textbf{(9) Robust Content Matching.}
  \else
    \item[(9)] \textbf{Robust Content Matching.}
  \fi
 Without a certificate ecosystem, there are still ways to identify media duplicated in unauthorized contexts. To build scanning systems that can help sex workers detect when their information has been shared (publicly) on the internet, we require a robust mechanism that can efficiently identify a sex worker's content.  
 Existing robust content matching techniques to match ``semantically equivalent'' images have been developed in the context of detecting child sexual abuse material (e.g., Microsoft's PhotoDNA, Facebook's PDQ, or Apple's NeuralHash).  The effectiveness of these techniques is questionable: the algorithms either require secrecy (PhotoDNA) or are more brittle than intended when made public (PDQ and NeuralHash).  In all cases, the efficacy of these algorithms requires that the media for which the scanner is searching remains secret; if the fingerprint of the media were revealed, it would be easy to modify the media such that it no longer matched.  It is not clear if existing approaches could be adapted to our setting, where different images could be semantically equivalent (e.g., different sex workers depicting the same activity) and fingerprints are not centrally curated.
    \ifdefined\maglayout

    \smallskip \noindent
    \textbf{(10) Anti-Theft Technology.}
  \else
    \item[(10)] \textbf{Anti-Theft Technology.}
  \fi
 It is not possible to completely prevent content theft---viewers can always take screenshots, photograph or record using another device, and most anti-pirating mechanisms can be bypassed. However, it is worth investing in technologies that make theft onerous. There are multiple ways to hinder downloading static content in HTML, and many commercial streaming services use anti-recording technology. Researchers could investigate how to make these solutions usable for small content-creators and small-scale software developers. Alternately, it may be possible to leverage cryptographic or steganographic watermarking to embed robust tags within the content could be used to identify its origin. While these techniques have been discussed in theory, to our knowledge they have not be meaningfully deployed or investigated from a systems perspective.
 \ifdefined\maglayout
 \else
\end{itemize}
\fi

\subsection{Relevance to Other Communities}
The directions we summarize are relevant not only to protecting sex workers, but to protecting other marginalized groups. Efforts around preventing or recovering from content theft in particular will benefit people participating in both sex work and recreational exchanges of intimate imagery. Technology that allows people to cryptographically prove ownership over images (8b) may be valuable for anyone who has their intimate images maliciously posted online. Systems for content matching (9) may help those who suspect that their intimate images were made public.

On the preventative side, people creating images for either commercial or recreational use may benefit from having anti-theft technology (10) and usable and effective image modification technology (5). For example, someone on a dating platform may be able to more comfortably engage in intimate exchanges if those photos are less likely to be screen shot and/or they are less likely to be recognizable in the photos. We note that both commercial and recreational creators may have complex use cases beyond those we explore here, such as creation of intimate media by or with another person, which complicates image ownership.

Other populations may also benefit from the lines of research we propose. For example, marginalized groups such as activists and racial and gender minorities also face de-platforming and heavy-handed content moderation. Experiments using filter analysis (1) and facial recognition circumvention (2) could be used to identify biases, and their causes, in content moderation against other groups. 
Further, tools that prevent outing like automated contact blocking (4) and multi-profile support (6) would be extremely valuable for other populations who frequently manage multiple, non-intersecting social spheres. For example, a transgender person who is in the process of coming out incrementally to friends and family needs to be able to control which self-presentation is visible to whom and in what context to stay safe. Finally, some directions, like the verification of privacy settings (7), stand to benefit all users.

\section{Conclusion}

As we demonstrate, there are many opportunities for systems security and cryptography research to support the safety of digital intimacy. As researchers embark on this work, we again offer several important considerations. First, as with all research on privacy, anonymity, and circumvention, some of this research may increase risks for some communities, potentially by aiding harmful behavior such as abuse.
Researchers must pay attention to the potential impact of new tools more broadly and work to minimize the risk of misuse. Second, fundamental problems that sex workers face---stigma, deliberate discrimination, and criminalization---and that those engaged in recreational sexual intimacy face---stigma, victim blaming---are rooted in misogyny and discrimination. Technology alone will not change the prevalence of these risks, only impact their online manifestations. 
Reducing the risks of digital intimacy requires substantive social change. 
Finally, and most importantly, centering affected communities directly in research conception and deployment is imperative to building systems that increase safety rather than harm. The people directly impacted by the products of the work will know best whether solutions reduce risk, increase risk, or merely rearrange it. There are a range of appropriate methods for centering marginalized users. Purely theoretical work (i.e., improving cryptographic primitives) should carefully leverage empirical work to appropriately ground theoretical use-cases, and applied work should engage directly with the people who may use---or be harmed by---the products of their work.

\bibliographystyle{plain}

\begin{thebibliography}{}

\end{thebibliography}


\begin{thebibliography}{10}

\bibitem{barakat2021community}
Hanna~L Barakat and Elissa~M Redmiles.
\newblock Community under surveillance: Impacts of marginalization on an online
  labor forum.
\newblock In {\em Proceedings of the International AAAI Conference on Web and
  Social Media}, 2021.

\bibitem{barwulor2021disadvantaged}
Catherine Barwulor, Allison McDonald, Eszter Hargittai, and Elissa~M Redmiles.
\newblock ``{Disadvantaged} in the {American}-dominated internet'': Sex, work,
  and technology.
\newblock In {\em Proceedings of the 2021 CHI Conference on Human Factors in
  Computing Systems}, 2021.

\bibitem{bhalerao2022ethics}
Rasika Bhalerao, Nora McDonald, Hanna Barakat, Vaughn Hamilton, Damon McCoy,
  and Elissa~M Redmiles.
\newblock Ethics and efficacy of unsolicited anti-trafficking {SMS} outreach.
\newblock {\em Proceedings of the ACM on Human-Computer Interaction}, (CSCW),
  2022.

\bibitem{blunt2020erased}
Danielle Blunt and Ariel Wolf.
\newblock Erased: The impact of {FOSTA-SESTA} and the removal of {Backpage} on
  sex workers.
\newblock {\em Anti-trafficking Review, 2020}, 2020.

\bibitem{blunt2020posting}
Danielle Blunt, Ariel Wolf, Emily Coombes, and Shanelle Mullin.
\newblock Posting into the void: Studying the impact of shadowbanning on sex
  workers and activists.
\newblock
  \url{https://hackinghustling.org/posting-into-the-void-content-moderation/},
  2020.

\bibitem{eaton20172017}
Asia Eaton, Holly Jacobs, and Yanet Ruvalcaba.
\newblock {\em 2017 Nationwide Online Study of Nonconsensual Porn Victimization
  and Perpetration}.
\newblock Cyber Civil Rights Initiative, Inc., 2017.

\bibitem{halperin2017war}
David~M Halperin and Trevor Hoppe.
\newblock {\em The war on sex}.
\newblock Duke University Press, North Carolina, United States, 2017.

\bibitem{hamilton2022risk}
Vaughn Hamilton, Hanna Barakat, and Elissa~M Redmiles.
\newblock Risk, resilience and reward: Impacts of shifting to digital sex work.
\newblock {\em Proceedings of the ACM on Human-Computer Interaction}, (CSCW),
  Forthcoming.

\bibitem{jones2020camming}
Angela Jones.
\newblock {\em Camming}.
\newblock New York University Press, 2020.

\bibitem{carceral_fosta}
Jody Liu.
\newblock The carceral feminism of {SESTA}-{FOSTA}: Reproducing spaces of
  exclusion from {IRL} to {URL}.
\newblock In {\em Queer Sites in Global Contexts: Technologies, Spaces, and
  Otherness, 2020}.

\bibitem{mcdonald2021s}
Allison McDonald, Catherine Barwulor, Michelle~L Mazurek, Florian Schaub, and
  Elissa~M Redmiles.
\newblock ``{It's} stressful having all these phones'': Investigating sex
  workers' safety goals, risks, and practices online.
\newblock In {\em {USENIX} Security 21}. USENIX Association, 2021.

\bibitem{sanders2018internet}
Teela Sanders, Jane Scoular, Rosie Campbell, Jane Pitcher, and Stewart
  Cunningham.
\newblock {\em Internet Sex Work: Beyond the Gaze}.
\newblock Springer, 2018.

\bibitem{smith2018revolting}
Molly Smith and Juno Mac.
\newblock {\em Revolting Prostitutes: The Fight for Sex Workers' Rights}.
\newblock Verso Trade, London, 2018.

\bibitem{spivsak2021social}
Sanna Spi{\v{s}}{\'a}k, Elina Pirjatanniemi, Tommi Paalanen, Susanna Paasonen,
  and Maria Vihlman.
\newblock Social networking sites’ gag order: Commercial content
  moderation’s adverse implications for fundamental sexual rights and
  wellbeing.
\newblock {\em Social Media+ Society}, 2021.

\bibitem{unaids}
UNAIDS.
\newblock Sex work and {HIV}/{AIDS}: {UNAIDS} technical update, 2002.

\end{thebibliography}

\begin{IEEEbiography}{Vaughn Hamilton}{\,} is a Research Associate at the Max Planck Insitute for Software Systems in Saarbrücken, Saarland 66123, Germany. Their research interests include technology use by marginalised groups, digital safety and ethics. Contact them at \texttt{vhamilto@mpi-sws.org}. 
\end{IEEEbiography}

\begin{IEEEbiography}{Gabriel Kaptchuk}{\,} is a Research Assistant Professor at Boston University, in Boston, Massachusetts, 02215, USA. His research interests include applied cryptography and privacy policy. Contact him at \texttt{kaptchuk.com}.
\end{IEEEbiography}

\begin{IEEEbiography}{Allison McDonald} is an Assistant Professor at Boston University in Boston, Massachusetts, 02215, USA. Her research interests include privacy and digital safety. Contact her at \texttt{amcdon.com}.
\end{IEEEbiography}

\begin{IEEEbiography}{Elissa M. Redmiles} is an Assistant Professor at Georgetown University in Washington, D.C., 20007, USA. 
Her research interests include security, privacy \& ethics.
Contact her at \texttt{elissaredmiles.com}.
\end{IEEEbiography}

\end{document}